\documentclass[onecolumn,showpacs]{revtex4}

\usepackage{amsmath}
\usepackage{makeidx}         
\usepackage{graphicx}        
\usepackage{epstopdf}
\usepackage{bm}

\makeindex             

\begin{document}

\title{Bose-Einstein condensates in 1D optical lattices: nonlinearity and Wannier-Stark spectra}
\author{Ennio Arimondo,  Donatella Ciampini and Oliver Morsch}
\affiliation{CNR-INFM and CNISM, Dipartimento di Fisica E.Fermi, Universit\`{a}
di Pisa, Via Buonarroti 2, I-56127 Pisa,Italy}

\begin{abstract}
We present our experimental investigations on the subject of
nonlinearity-modified Bloch-oscillations and of nonlinear Landau-Zener
tunneling between two energy bands in a rubidium Bose Einstein condensate in an
accelerated periodic potential.
Nonlinearity introduces an asymmetry in Landau-Zener tunneling.  We also present
measurements of resonantly enhanced tunneling between the Wannier-Stark energy levels for
Bose-Einstein condensates  loaded into an optical lattice.
\end{abstract}

\maketitle

\section{Introduction}
The development of powerful laser cooling and trapping techniques has made possible the controlled realization of dense and cold gaseous samples, thus opening the way for investigations in the ultracold temperature regimes not accessible with conventional techniques. A Bose-Einstein condensate (BEC) represents a peculiar gaseous state where all the particles reside in the same quantum mechanical state.  Therefore BEC's exhibit quantum mechanical phenomena on a macroscopic scale with a single quantum mechanical wavefunction describing the external degrees of freedoms. That  control of the external degrees of freedom is  combined with a precise control of the internal degrees.  The BEC investigation has become a very active area of research in contemporary physics. The BEC study  encompasses different subfields of physics, i.e., atomic and molecular physics, quantum optics, laser spectroscopy, solid state physics. Atomic physics and laser spectroscopy provide the methods for creating and manipulating the atomic and molecular BECÕs. However owing to the interactions between the particles composing the condensate and to the configuration of the external potential,  concepts and methods from solid state physics are extensively used for BEC description.

Quantum mechanical BEC's within the periodic potential created by interfering laser waves ("optical lattices")  have attracted a strongly increasing interest\cite{Morsch02,Bloch05a,Bloch05b,Morsch06}. In particular, the formal similarity between the
wavefunction of a BEC inside the periodic potential of an optical lattice and
electrons in a crystal lattice has triggered theoretical and experimental efforts
alike. BEC's inside optical lattices share many features with electrons in solids, but also with light waves in nonlinear materials and other nonlinear systems.  However,
the experimental control over the parameters of BEC and of the periodic potential make it
possible to enter regimes inaccessible in other systems.
 Many phenomena from condensed matter physics, such as Bloch oscillations and
Landau-Zener tunneling have since been shown to be observable also in optical
lattices.  BEC in an
optical lattice even made possible the observation of a quantum phase transition
that had, up to then, only been theoretically predicted for condensed matter
systems~\cite{Greiner02}. However an important difference between electrons in a
crystal lattice and a BEC inside the periodic potential of an optical lattice is
the strength of the self interaction between the BEC components and hence the magnitude of the nonlinearity of the system. Electrons are almost noninteracting whereas atoms inside a BEC interact
strongly. A perturbation approach is appropriate in the former case while in the
latter the full nonlinearity must be taken into account. Generally, atom-atom
interactions in Bose-Einstein condensates lead to rich and interesting nonlinear effects.  Most experiments to date have been carried out in the regime
of shallow lattice depth, for which the system is well described by the
Gross-Pitaevskii equation, a mean-field equation.
Moreover, the nonlinearity induced by the mean-field of the condensate has been
shown both theoretically and experimentally to give rise to
instabilities in certain regions of the Brillouin zone. These instabilities are not present in
the corresponding linear system, i.e. the electron system.

The present text initially describes the construction of the optical lattice periodic potential for cold atoms in Sec. \ref{lattice}. The following Section reports the analysis of the condensate interference pattern when released from the optical lattice.  In Sec. \ref{nonlinear} the nonlinear term within the Gross-Pitaevskii equation describing the dynamics of a Bose-Einstein condensation is introduced. The following Sections report experimental results on the Bloch oscillations, on the nonlinear Landau-Zener quantum tunneling and on the resonantly enhanced quantum tunneling. A short conclusion terminates the presentation.

\section{Optical lattice}
\label{lattice}
In order to trap a Bose-Einstein condensate in a periodic potential, it is sufficient to
exploit the interference pattern created by two or more
overlapping laser beams and the light force exerted on
the condensate atoms. Optical lattices  work on the
principle of the ac Stark shift. When an atom is placed in
a light field, the oscillating electric field of the latter induces
an electric dipole moment in the atom. The interaction
between this induced dipole and the electric field
leads to an energy shift $\Delta E$ of an atomic energy level. When we take two
identical laser beams  and make them
counterpropagate in such a way that their cross sections
overlap completely see Fig. 1(a),  we
expect the two beams to create an interference pattern,
with a distance $d_{\rm L}=\lambda/2$ between two neighbooring maxima or minima of
the resulting light intensity. The potential seen by the
atoms is then
\begin{equation}
V(x) = V_0\cos^2\left (\frac{\pi x}{d}\right)= \frac{V_0}{2}\left[1+\cos\left (\frac{2\pi x}{d_{\rm L}}\right)\right],
\end{equation}
where the lattice
depth $V_0$ is determined by the light shifts $\Delta E$ produced by the individual laser beams.
 The easiest option to create a one-dimensional optical lattice is to take a linearly polarized laser beam and retro-reflect it with a
high-quality mirror. If the retro-reflected beam is replaced by a second
phase-coherent laser beam as obtained, for
instance, by dividing a laser beam in two, another degree of freedom is introduced. It is now possible to have a frequency shift $\Delta \nu_{\rm L}$ between the two lattice beams. The periodic lattice
potential will now no longer be stationary in space but move at a
velocity $v_{\rm lat} = d_{\rm L}\Delta \nu_{\rm L}.$
If the frequency difference is varied at a rate $\partial\Delta  \nu_{\rm L}/ \partial t$,
the lattice potential will be accelerated with $a_{\rm lat} = d_{\rm L} \partial \Delta \nu_{\rm L}/ \partial t.$
Clearly, in the rest frame of the lattice there will be a
force
\begin{equation}
F=-Ma_{\rm lat}=-Md_{\rm L} \frac{\partial \Delta \nu_{\rm L}}{\partial t}
\end{equation}
 acting on the condensate atoms. This gives us a powerful tool for manipulating
a BEC inside an optical lattice.

\begin{figure}[htbp]
\includegraphics[scale=0.25]{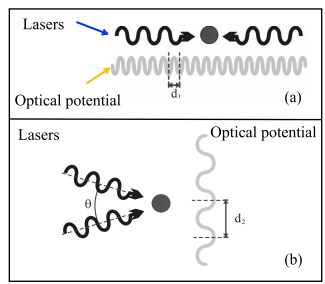}
 \caption{Representation of the laser configuration creating an optical lattice in the counterpropagating geometry, in (a), and in the angle tuned geometry, in (b)}
\label{LatticeConf}
\end{figure}

Another degree of freedom of a 1D lattice realized
with two laser beams is the lattice constant. The spacing
$d_{\rm L}$ between two adjacent wells of a lattice resulting
from two counterpropagating beams can be enhanced by
making the beams intersect at an angle $\theta \ne  \pi$, see Fig.
1(b). This
will give rise to a periodic potential with lattice constant
\begin{equation}
d_{\rm L}=\frac{\lambda}{2\sin(\theta/2)}.
\end{equation}
 To simplify the notation, in the following
we shall always denote the lattice constant by $d_{\rm L}$ and all
the quantities derived from it, regardless
of the lattice geometry that was used to achieve it. In particular the recoil energy
 $E_{\rm R}$ and the recoil frequency $\nu_{\rm R}$ of an atom with mass $M$ are
 \begin{equation}
  E_{\rm R}=h \nu_{\rm R}= \frac{\hbar^2 \pi^2}{2 Md_{\rm L}^2},
  \end{equation}
and the recoil velocity $v_{\rm R}=\hbar \pi/(d_{\rm L}M)$. Naturally, by adding more laser beams one can
easily create two- or three-dimensional lattices\cite{Morsch06}.

\begin{figure}[htbp]
\includegraphics[scale=0.8]{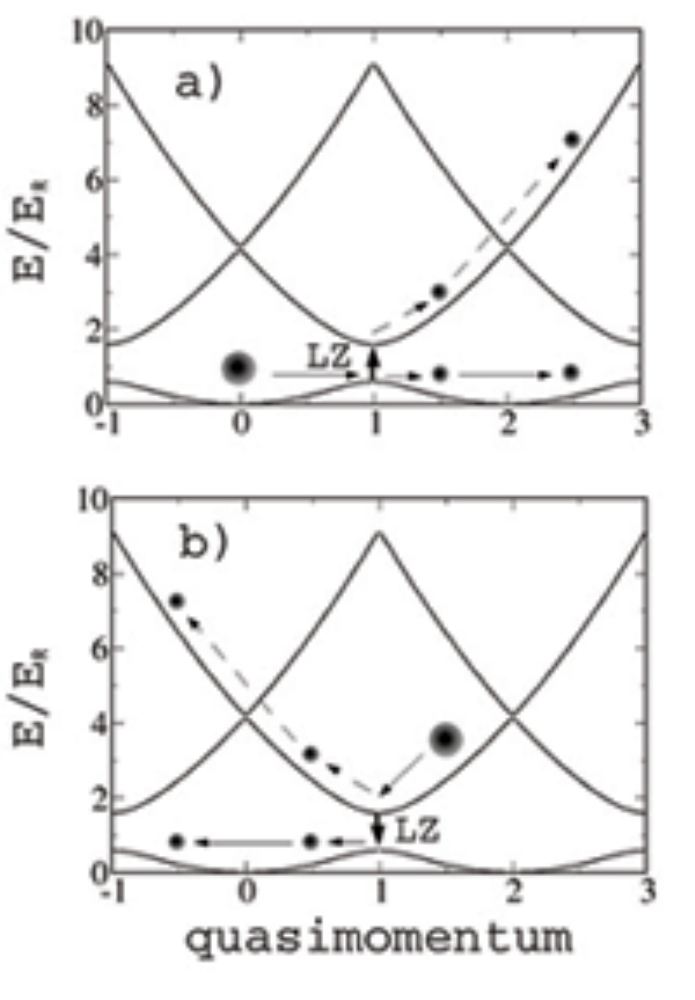}
 \caption{Band structure of a BEC in an
optical lattice ($V_0=2\,E_{\rm R}$) and LZ-tunneling (ground to
excited state (a) and excited to ground state (b)). When the BEC
is accelerated across the edge of the Brillouin zone (BZ) at
quasimomentum $q/q_{\rm R}=1$, LZ tunneling occurs.  After the first crossing of the edge of
the Brillouin zone increasing the lattice depth and
decreasing the
acceleration leads to a much reduced tunneling rate from the
ground state band at successive BZ-edge crossings, as shown in (a)
for the ground to excited state tunneling and in (b) for excited
to ground state}
\label{Bandstructure}
\end{figure}

The description of the propagation of noninteracting
matter waves in periodic potentials is straightforward
once one has found the eigenstates and corresponding
eigenenergies of the system. The eigenstates are found in by
applying Bloch's theorem, which states that the eigenfunctions have the form \cite{AshcroftMermin76}
\begin{equation}
\phi_{\rm n,q}(x) = e^{iqx}u_{\rm n,q}(x),
\end{equation}
where $\hbar q$ is referred to as quasimomentum and $n$ indicates
the band index, the meaning of which will become
clear in the following discussion. The quasimomentum $q$ appearing in the Bloch's theorem can always be confined to the first Brillouin zone $(-q_{\rm R},q_{\rm R})$ with $q_{\rm R}=\pi/d_{\rm L}$ , because any $q^\prime$ not in the first Brillouin zone can be written as $q^\prime = q+G$, where $G$ is a reciprocal lattice vector and $q$ does lie in the first zone.
The eigenenergies $E^n(q)$ of the above eigenstates depend on the potential
depth $V_0=sE_{\rm R}$ and, additionally, on the quasimomentum
$q$. In Fig. \ref{Bandstructure}, we summarize the properties of the
eigenbasis for a shallow potential $V_0=2E_{\rm R}$. The eigenenergies form bands that are
separated by a gap in the energy spectrum, i.e., certain
energies are not allowed. Since
the gap energy $E_{\rm gap}^n$
between the $n$th and $(n+1)$ th band scales with $V_0^{
n+1}$ in the weak potential
limit, it only has appreciable magnitude
between the lowest and first excited band. A
particle with high energy is very well described as a free
particle and the influence of the periodic potential is
negligible in this case.
It is important to note that for energies
near the Brillouin zone edge of the lowest band, the
eigenstate probability distribution is a periodic $\sqrt{2}\sin\frac{2 \pi x}{2d_{\rm L}}e^{i\frac{\pi x}{2d}}$ function, its maxima coinciding
with the potential minima and the phase
changing by $\pi$ between adjacent wells.  This is
the well-known sinusoidal Bloch state at the
Brillouin zone edge, in the literature also referred to as a Òstaggered mode.
In the limit of deep periodic potentials, also referred to
as the tight-binding limit, the eigenenergies of the low lying
bands are only weakly dependent on the quasimomentum.

The wave-packet dynamics of a particle  in a periodic potential in
the presence of an additional external potential, i.e.,
with an external force, is generally not easy to solve. The
problem becomes relatively simple, though, as soon as
the width of the wave packet in quasimomentum space
is small and thus the wave packet can be characterized
by a single mean quasimomentum $q(t)$ at time $t$. An external
force then leads to a time-dependent $q(t)$ via
\begin{equation}
\hbar {\dot q}(t)=F.
\end{equation}
 In
the case of a constant force $F$ e.g., due to the gravitational
field, this results in
\begin{equation}
q(t)=q(t=0)+\frac{Ft}{M}.
\end{equation}
In addition, the velocity $v_{\rm n}(q)$ of the particle  in the $n$ band is given by the group velocity of the underlying
wavepacket
 \begin{equation}
v_{\rm n}(q)=\frac{1}{\hbar}\frac{\partial E_{\rm n}(q)}{\partial q}.
\end{equation}
The above equations determine that the rate of change of the quasimomentum is given by the external force, but the rate of change of the wavepacket's momentum
is given by the total force including the influence of the periodic field of the lattice.
In the case of a constant force, the velocity at time $t$ is
\begin{equation}
v_{\rm n}\left(q(t)\right)=v_{\rm n}\left(q(t=0)\right)+\frac{Ft}{M}.
\end{equation}
Since $v_{\rm n}$ is periodic in the reciprocal lattice, the velocity is a bounded and oscillatory function of time. Therefore the result of the force is not an acceleration of
the wave packet, and instead the wavepacket will show an oscillatory
behavior in real space.  The velocity oscillatory motion is
known as Bloch oscillations~\cite{Bloch29} and the period as the Bloch time
\begin{equation}
T_{\rm B}=\frac{2\pi \hbar}{Fd_{\rm L}}=\frac{1}{F_0 \nu_{\rm R}},
\label{Blochperiod}
\end{equation}
where we have introduced a dimensionless force $F_0=F d_{\rm L}/E_{\rm R}$.

In the case of a strong external force acting on matter
waves in periodic potentials, transitions into higher
bands can occur as schematically represented in Fig. \ref{Bandstructure}. In the context of electrons
in solids, this is known as the Landau-Zener (LZ) breakdown~\cite{Landau32,Zener32},
occurring if the applied electric field is strong enough for
the acceleration of the electrons to overcome the gap
energy separating the valence and conduction bands.
It was shown in~\cite{Zener32} that for a given acceleration
$a_{\rm L}$ corresponding to a constant force, one can
deduce a tunneling probability across the first-second band gap in the adiabatic limit
\begin{equation}
P_{\rm LZ} =e^{-\frac{\pi^2V_0^2}{32F_0}}.
\label{landauzener}
\end{equation}
The resulting
wavepacket dynamics is shown in Fig. \ref{Bandstructure}, where LZ
tunneling, from $n=1$ to $n=2$ band, leads to a splitting of the wave function.

\section{Analysis of the interference pattern}
\label{analysis}
\begin{figure}[htbp]
\includegraphics[scale=0.5]{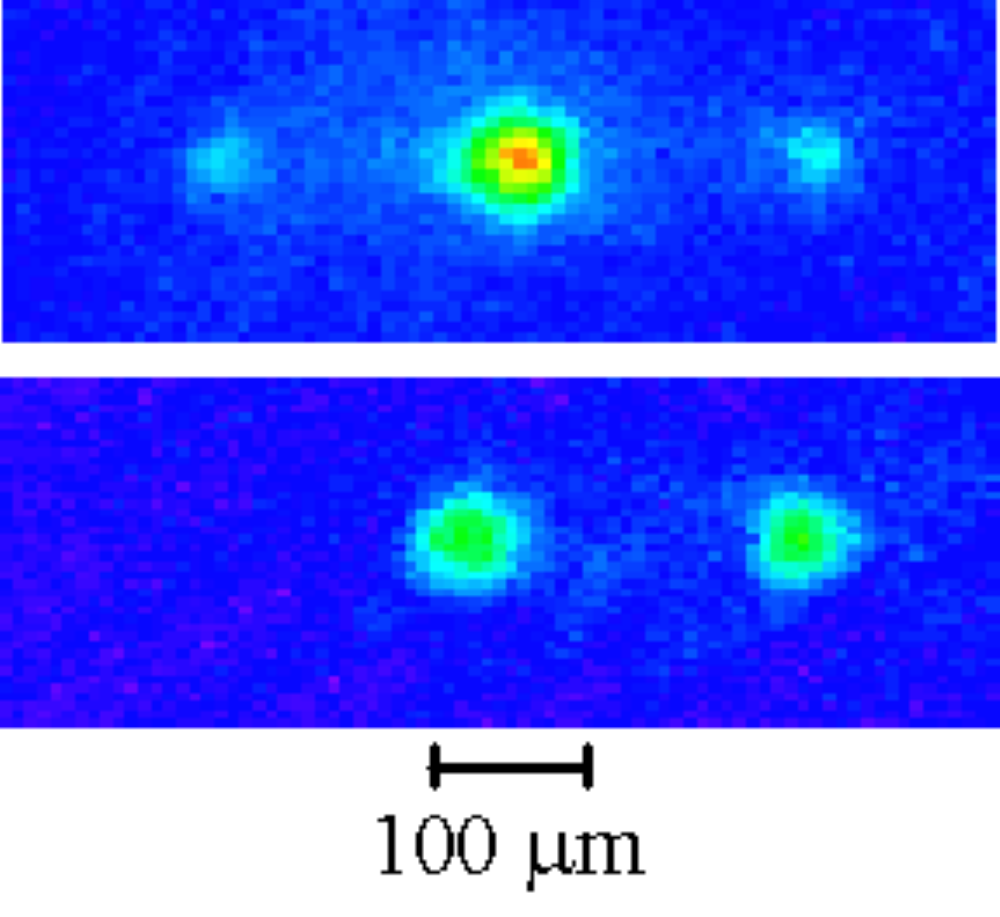}
\caption{Interference pattern of a Bose-Einstein condensate
released from a one-dimensional optical lattice of
depth $V_0=10E_{\rm R}$ after a time of flight of 20 ms. In the top pattern the lattice
was at rest, whereas in the bottom one the condensate  had been accelerated to $v_{\rm R}$, i.e.,
its quasimomentum was at the edge of the
Brillouin zone}
\label{CondensImage}
\end{figure}

 Doing experiments with condensates in optical lattices
is useful only if one is able to extract information from
the system once the experiment has been carried out. There are essentially
two methods for retrieving information from
the condensate: in situ and after a time of flight. In the
former case, one can obtain information about the spatial density
distribution of the condensate, its shape, and
any irregularities on it that may have developed during the
interaction with the lattice. Also, the position of the center
of mass of the condensate can be determined.
Looking at a condensate released from a lattice after a time of flight, typically on the order of a few milliseconds,
 amounts to observing its momentum
distribution. When the atomic system is in a steady state,
   the condensate is distributed among the lattice wells (in the limit of
a sufficiently deep lattice in order for individual lattice sites to have well-localized wavepackets). If the lattice is now switched off suddenly, the individual
   (approximately) Gaussian wavepackets at each lattice site will
   expand freely and interfere with one another. The resulting
   spatial interference pattern after a time-of-flight of $t$ will be
   a series of regularly spaced peaks with spacing $2v_{\rm R} t$,
corresponding to the various diffraction orders. In the case of a condensate that is very elongated along the lattice
direction, to a good approximation we initially
have an array of equally spaced Gaussians of a
width  d determined by the lattice depth.
Figure \ref{CondensImage} shows a typical time-of-flight interference
pattern of a condensate released from an optical lattice
(plus harmonic trap)  for a lattice depth $V_0=10E_{\rm R}$. From
the spacing of the interference peaks and the time of
flight, one can immediately infer the recoil momentum
of the lattice and hence the lattice constant d. Furthermore,
from the relative height of the side peaks corresponding
to the momentum classes $\pm 2 \pi/ d_{\rm L}$, one can calculate
the lattice depth. The top interference pattern was produced by a condensate at rest
with zero quasimomentum. Instead the bottom interference pattern was produced by a condensate with
quasimomentum at the edge of the Brillouin zone, in the staggered state with the condensate  wavefunction  changing by $\pi$ between adjacent wells.

\section{Nonlinear optical lattice}
\label{nonlinear}
\indent The motion of a Bose-Einstein
condensate in a 1D optical lattice experiencing an acceleration $a_{\rm L}$ is described by the Gross-Pitaevskii
equation
\begin{equation}
 \label{schrod}
i\hbar \frac{\partial \psi}{\partial t} =
\frac{1}{2M}\left(-i\hbar\frac{\partial}{\partial
x}+Ma_{L}t\right)^2\psi +   \frac{V_0}{2}\cos\left(\frac{\pi x}{d_{\rm L}}\right)\psi +
\frac{4\pi\hbar^2a_s}{M}\left|\psi\right|^2\psi.
\end{equation}
The $s$-wave scattering
length $a_s$  determines the nonlinearity of the system. Equation
\ref{schrod} is written in the comoving frame of the lattice, so
the inertial force $-Ma_{\rm L}$ appears as a momentum modification. The
wavefunction $\psi$ is normalized to the total number of atoms in
the condensate and we define $n_0$ as the average uniform atomic
density. Defining the dimensionless quantities $\tilde x=2\pi x/d_{\rm L}$, $\tilde t=8E_{\rm R}t/\hbar$, and
rewriting $\tilde{\psi} =\psi/\sqrt{n_0}$, $\tilde
v=V_0/16E_{\rm R}$, $\tilde \alpha=M a_Ld_{\rm L}/16E_{\rm R}\pi$, $C= a_s
n_0d_{\rm L}^2/\pi$,  \ref{schrod} is cast in the following
form:
\begin{equation}
 \label{schrod-adim}
i\frac{\partial \psi}{\partial t} = \frac{1}{2}\left(
-i\frac{\partial}{\partial x}+\alpha t \right)^2 \psi +v
\cos(x)\psi + C \left|\psi \right|^2 \psi,
\end{equation}
where we have replaced $\tilde x$ with $x$, etc. In the
neighborhood of the Brillouin zone edge we can approximate the
wave function by a superposition of two plane waves, assuming that only the ground
state and the first excited state are populated. We then
substitute in  \ref{schrod-adim} a normalized wavefunction
\begin{equation}
    \psi(x,t)=a(t)e^{iqx}+b(t)e^{i(q-1)x}.
\end{equation}

Projecting on this basis,  linearizing
   the kinetic terms and dropping the irrelevant constant energy,  \ref{schrod-adim} assumes the form
\begin{equation}
\label{two-state-eq}
i\frac{\partial}{\partial t}\,\begin{pmatrix} a \\ b \end{pmatrix} =
     \left[-\frac{\alpha t}{2}\sigma_{3}+\frac{v}{2}\sigma_{1}\right]
     \begin{pmatrix} a \\ b \end{pmatrix}
+\frac{C}{2}(|b|^2-|a|^2)\sigma_{3} \begin{pmatrix} a \\ b \end{pmatrix},
\end{equation}
where $\sigma_i\,\,i=1,2,3$ are the Pauli matrices.  The adiabatic
energies of \ref{two-state-eq} have a butterfly structure at the
band edge of the Brillouin zone for $C \ge
v$~\cite{Wu03}, but in the present work we only
consider a regime where $C\ll v$, hence that structure plays no role.

In the linear regime $(C=0)$, evaluating the transition probability
in the adiabatic approximation, we find the linear LZ formula for the
tunneling probability $P_{\rm LZ}$ given
by \ref{landauzener}. Therefore for $C=0$ the tunneling
probability is the same for both tunneling directions whereas for $C\neq
0$ the two rates are different.  In the
nonlinear regime, as the nonlinear parameter $C$ grows, the lower
to upper tunneling probability grows as well until an adiabaticity
breakdown occurs at  $C=v$~\cite{Wu03}. The upper to lower
tunneling probability, on the other hand, decreases with
increasing nonlinearity.

The asymmetry in the tunneling transition probabilities can be
explained qualitatively as follows: The nonlinear term of the
Schr\"odinger equation acts as a perturbation whose strength is
proportional to the energy level occupation. If the initial state
of the condensate in the lattice corresponds to a filled lower
level of the state model, then the lower level is shifted upward
in energy while the upper level is left unaffected. This reduces
the energy gap between the lower and upper level and enhances the
tunneling. On the contrary, if all atoms fill the upper level then
the energy of the upper level is increased while the lower level
remains unaffected. This enhances the energy gap and reduces the
tunneling. The overall balance leads to an asymmetry between the
two tunneling processes.

The nonlinear regime may be reinterpreted by writing \ref{two-state-eq} as
\begin{equation}
 \label{two-state1-eq}
i\frac{\partial}{\partial t}\,\begin{pmatrix} a \\ b \end{pmatrix}
=  \left[-\frac{\alpha t}{2}\sigma_{3}+\frac{v}{2}\sigma_{1}\right]
     \begin{pmatrix} a \\ b \end{pmatrix}
- \frac{C}{2} \begin{pmatrix} |a|^2 & -b^*a \\ -a^*b & |b|^2
\end{pmatrix} \begin{pmatrix} a \\ b \end{pmatrix}.
\end{equation}
The nonlinear off-diagonal terms modify the interaction
term $v$ in a way equivalent to a Rabi frequency in the two-level
model.
In \ref{two-state1-eq} we identify
an offdiagonal term $v+C\,a^*b$ which acts as an effective potential.
Thus for small $C$ values we can modify the linear LZ formula \ref{landauzener}
to include nonlinear corrections, substituting the potential $v=V_0/16 E_{\rm R}$ with the
effective potential $v_{\rm eff}=V_{\rm eff}/16E_{\rm R}\equiv |v+C\,a^*b|$ (modulus is needed since
$a^*b$ is complex).
The expression for $v_{\rm eff}$ is
\begin{equation}
v_{\rm eff}=v\sqrt{1\pm\frac{C}{v}+\frac{C^2}{4v^2}},
\end{equation}
where the upper and lower signs corresponds to initial conditions
of excited/ground states.

\section{Bloch oscillations}
\label{oscillations}
A fundamental property of a quasiparticle in a periodic potential subject to an external static force is its localization by Bragg reflections at the boundary of the Brilouin zone, which leads to temporal and spatial oscillations known as Bloch oscillations~\cite{Bloch29}.  Related fundamental transport phenomena are the nonresonant LZ tunneling into a continuum of states of another Bloch band and the resonant LZ tunneling between anticrossing Wannier-Stark states of neighboring Bloch bands.
Bloch oscillations were first observed as time-resolved oscillations of wave packets of photo-excited "hot" electrons in biased semiconductor superlattices. Later Bloch oscillations and LZ tunneling were observed in ensembles of cold atoms~\cite{BenDahan96,Peik97a,Peik97b}.
During the last decade, there were experimentally realized time-resolved Bloch oscillations of coherent electron wave packets in semiconductor superlattices~\cite{Waschke93,Lyssenko97} subjected to combined electric and magnetic fields.  The progress in the
fabrication and investigation of complex optical nanostructures has allowed for direct experimental observations of one-dimensional optical Bloch oscillations of an optical laser field
in dielectric structures with a transversely
superimposed linear ramp of the refractive index~\cite{ Pertsch99,Morandotti99}. A
periodic distribution of the refractive index plays a role of
the crystalline potential, and the index gradient acts similarly
to an external force in a quantum system.  This force  causes the laser
beam to move across the structure while experiencing
Bragg reflections on the high-index and total internal reflection
on the low-index side of the structure, resulting in
an optical analogue of Bloch oscillations.
Bloch oscillations and LZ  tunneling from the first to second energy band was also demonstrated experimentally in two-dimensional photonic structures~\cite{Trompeter06}.   Most recently, acoustic Bloch oscillations and resonant LZ  tunneling of phononic wave packets were observed in perturbed ultrasonic superlattices~\cite{SanchisAlepuz07}.

\begin{figure}
\begin{center}
\includegraphics[scale=0.4]{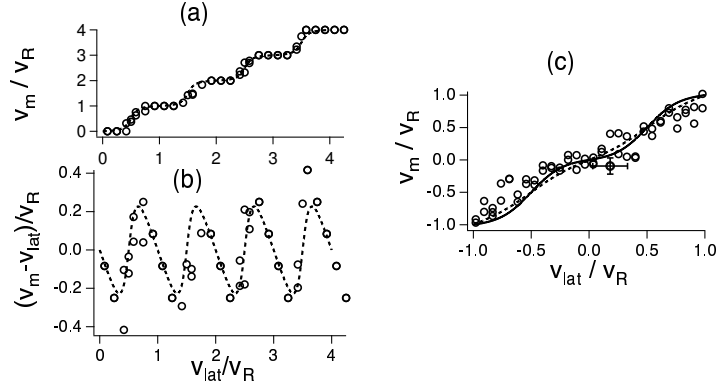}
\end{center}
\caption{Bloch oscillations of a Bose-Einstein condensate in an optical lattice. (a) Acceleration
in the counterpropagating lattice with $d_{\rm L}=390$ nm, $V_0\approx
2.3\,E_{\rm R}$ and
$a=9.81\,\mathrm{m\,s^{-2}}$. Dashed line: theory. (b) Bloch oscillations in the rest frame of the
lattice, along with the theoretical prediction (dashed line) derived from the shape of the lowest
Bloch band. (c) Acceleration in a lattice with $d_{\rm L}=1.56\,\mathrm{\mu m}$ and $V_0\approx
11\,E_{\rm R}$. In this case,the Bloch oscillations are much less pronounced. Dashed and solid lines:
theory for $V_0=11\,E_{\rm R}$ and $V_{eff}\approx7\,E_{\rm R}$}
\label{fig:blochoscillations}
\end{figure}

We report here results for Bloch oscillations in experiments with Bose-Einstein
   condensates adiabatically loaded into one-dimensional optical
   lattices. In particular,  we discuss
the dynamics of the BEC
   when the periodic potential provided by the optical lattice is
   accelerated, leading to Bloch oscillations ~\cite{Morsch01,Cristiani02}.
  The condensate was loaded adiabatically into the (horizontal)
optical lattice with lattice constant $d_{\rm L}=390$  nm
   immediately after switching off the magnetic trap. Thereafter, the lattice
   was accelerated with $a=9.81\,\mathrm{m\, s^{-1}}$ by ramping the
   frequency difference $\Delta \nu_{\rm L}$ between the laser beams forming the optical lattice. After a time
    the lattice was switched off and the condensate was
   observed after an additional time-of-flight. Fig.~\ref{fig:blochoscillations}(a) and (c) shows the results
   of these measurements in the laboratory frame. The Bloch
   oscillations are more evident, however, if one calculates the
   mean velocity $v_{\rm m}$ of the condensate as the weighted sum over the momentum components
   after the interaction with the accelerated lattice, as shown in
   Fig.~\ref{fig:blochoscillations}(b). When the instantaneous lattice
   velocity $v_{\rm lat}$ is subtracted from $v_{\rm m}$, one clearly sees the
   oscillatory behaviour of $v_m-v_{\rm lat}$.  The added feature of using a Bose-Einstein condensate is that    the spatial extent of the atomic cloud is sufficiently small so
that after a relatively short time-of-flight the separation between the
individual momentum classes is already
   much larger than the size of the condensate due to its expansion
and can, therefore,  be easily resolved. Similar observations were reported in \cite{HeckerDenschlag02,Browaeys05}. In ref. \cite{Schmid06}  by using an optical Bessel beam to form the optical lattice, a very large number of Bloch oscillations of a rubidium condensate was realized and large  final velocities were reached.

\begin{figure}[htbp]
\includegraphics[scale=1]{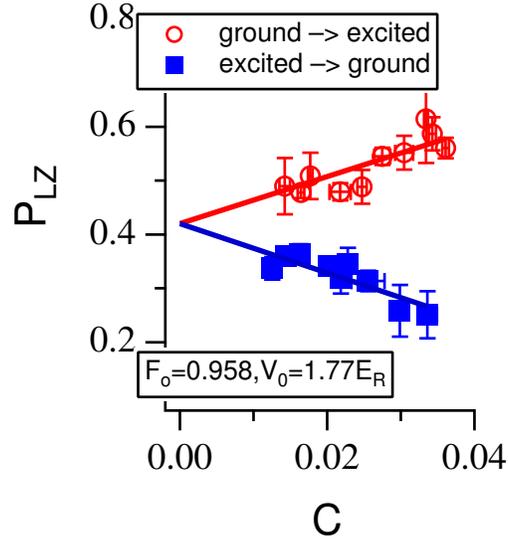}
\caption{Asymmetric tunneling between the ground state and first
excited band of a BEC in an optical lattice as a function of the
nonlinear interaction parameter $C$.
In these experiments, $a=32.1\,\mathrm{m\,s^{-2}}$ and the lattice depth
was $1.77\,\mathrm{E_{\rm R}}$}
\label{Fig_twodir}
\end{figure}

\section{Landau-Zener tunneling}
\label{LandauZener}
The linear regime of the LZ  tunneling in atomic physics  was investigated by several authors,  using Rydberg atoms in \cite{Rubbmark81}, in classical optical systems in \cite{Bouwmeester95}, for cold atoms in an accelerated optical potential in \cite{Bharucha97}. We investigated linear and nonlinear LZ tunneling between the two lowest energy bands of a
condensate inside an optical lattice
   in the following way.
   Initially, the condensate was loaded adiabatically into
   one of the two bands.
   Subsequently, the lattice was accelerated in such a way that
   the condensate crossed the edge of the Brillouin zone once,
   resulting in a finite probability for tunneling into the other
   band (higher-lying bands can be safely neglected as their
   energy separation is much larger than the band gap). Thus, the two bands had populations reflecting the
   LZ  tunneling probability (assuming only one band exclusively populated initially). In order to
   experimentally determine the number of atoms in the two bands, we
   then {\em increased} the lattice depth  and {\em decreased} the
acceleration.
   Using this experimental sequence we selectively accelerated further
   that part of the condensate  that populated the ground
   state band, whereas the population of the first excited band was
   not accelerated further, as shown schematically in Fig. \ref{Bandstructure}.

 In order to investigate tunneling from the ground state band to
   the first excited band, we adiabatically ramped up the lattice
   depth with the lattice at rest and then started the
   acceleration sequence. The tunneling from the first excited to the
ground-state band
   was investigated in a similar way, except that in this case we
   initially prepared the condensate in the first excited band by
   moving the lattice with a velocity of $1.5\,v_{\rm R}$ (through the
frequency difference $\Delta \nu_{\rm L}$ between the acousto-optic
   modulators) when
   switching it on. In this way, in order to conserve energy and
   momentum the condensate must populate the first excited band at
   a quasi-momentum half-way between zero and the edge of the
   first Brillouin zone. For both tunneling
   directions, the tunneling probability is derived from
   $P_{\rm LZ}=N_{\rm tunnel}/N_{\rm tot},$
     where $N_{\rm tot}$ is the total number of atoms measured from the
   absorption picture. For the tunneling from the first
   excited band to the ground band, $N_{\rm tunnel}$ is the number of
   atoms accelerated by the lattice,  whereas for the inverse
   tunneling direction, $N_{\rm tunnel}$ is the number of atoms
   detected in the $v=0$ velocity class.

For a small value of the interaction
parameter $C$, we verified that the tunneling rates in the two directions are
essentially the same and agree well with the linear LZ
prediction.  By contrast, when $C$ is increased, the two tunneling
rates differ greatly, as in Fig.~\ref{Fig_twodir}.
We have not yet performed a quantitative comparison of these data with the theoretical predictions
of the non-linear LZ  model. Previous data published in \cite{Jona03} presented good agreement with the theoretical predictions of the asymmetric  nonlinear LZ tunneling.

\begin{figure}[htbp]
\includegraphics[scale=0.3]{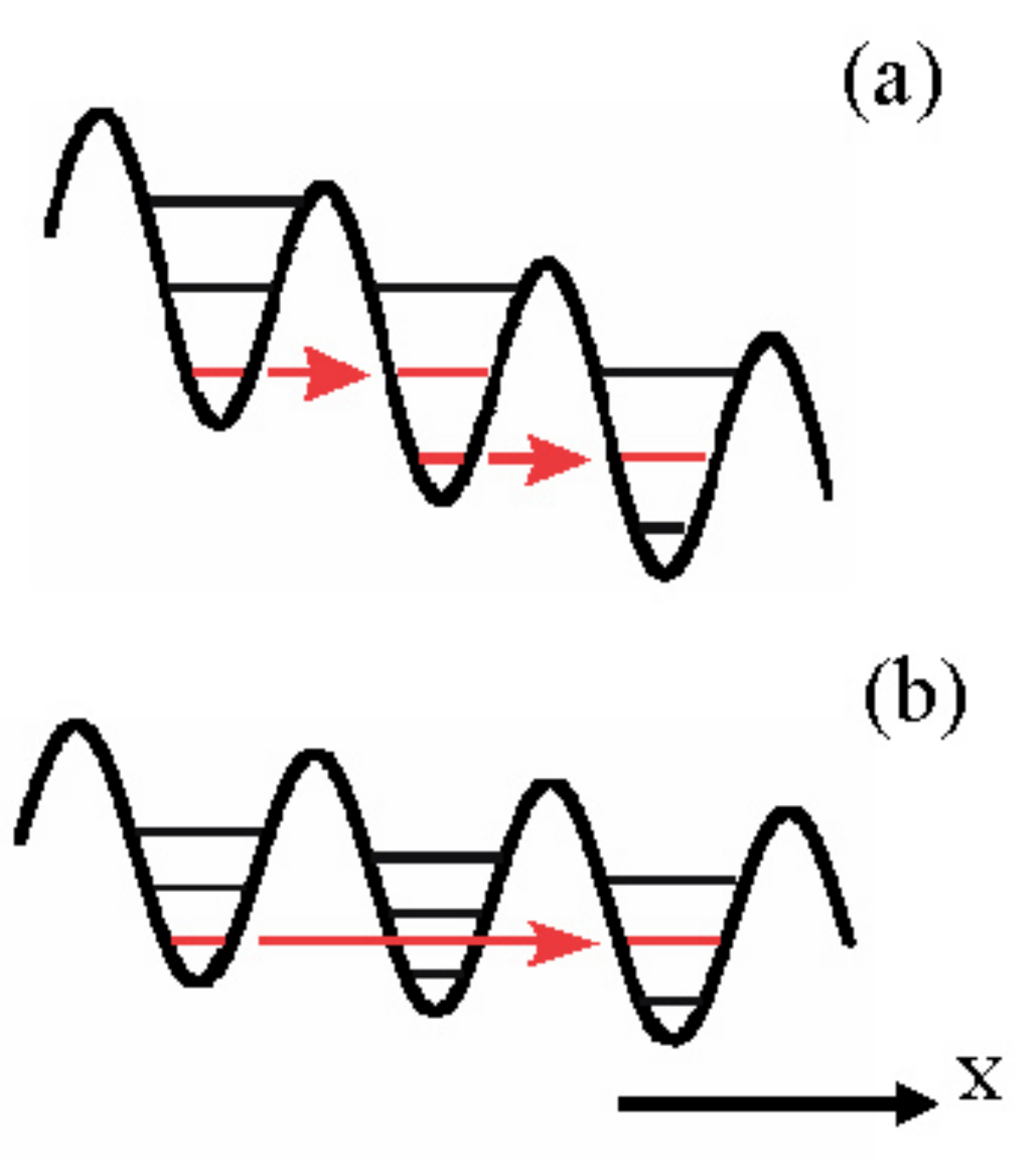}
\caption{The tunneling of atoms out of a tilted lattice is
resonantly enhanced when the tilt induced energy difference $Fd_L\Delta i$ between lattice
wells $i$ and $i+\Delta i$ matches the separation between two quantized energy levels within a well, $\Delta i=1$ in the top and $\Delta i=2$ in the bottom}
\label{WannierStarkConf}
\end{figure}

\section{Resonantly enhanced quantum tunnelling}
\label{RETSection}
Resonantly enhanced tunneling (RET) is a quantum effect in which
the probability for tunneling of a particle between two potential
wells is increased when the quantized energies of the initial and
final states of the process coincide. In spite of the fundamental
nature of this effect and the practical
interest, it has been difficult to observe
experimentally in solid state structures. Quantum tunnelling has found many technological applications,  for instance, in
scanning tunnelling microscopes and in superconducting squid devices. The
most widely application is in tunnelling diodes and related integrated
semiconductor devices which go back to the pioneering work of Leo Esaki~\cite{Esaki73}. The latter also
proposed to exploit resonantly enhanced tunnelling (RET) for technical use, and since the 1970's
much progress has been made in producing artificial superlattice structures, in which RET
of fermionic quasiparticles could be demonstrated.

Here we present our realization of RET using Bose-Einstein condensates  held in
optically induced potentials. The counter-propagating beams creating the
lattice were continuously accelerated such as to
mimic a static linear potential in the moving frame of reference. BEC tunneling
occurred between the quantised energy levels (theWannier-Stark levels)
in various wells of the potential, see Fig. \ref{WannierStarkConf}.  We demonstrated that the tunneling probability is resonantly
enhanced and the LZ formula does not give the correct
result.
\begin{figure}[htbp]
\begin{center}
\includegraphics[scale=0.5]{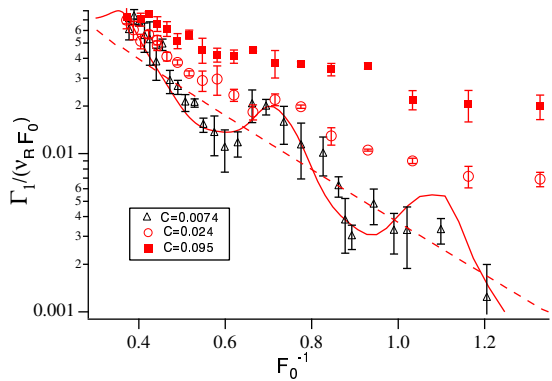}
\end{center}
\caption{ Tunneling resonances of the $n=1$ lowest energy
level for $V_0=3.5\,E_{\rm R}$. The continuous line represents the theoretical RET
prediction in the linear regime, and  the dashed line the
LZ  prediction. For the experimental data (at $d_{\rm L}=426.1$ nm and $C=0.007$) and theoretical prediction in the  RET linear regime, the $\Delta i=2$ and $\Delta i=3$ resonant peaks appear at increasing values of $F_0^{-1}$, while the $\Delta i=1$
peak is not completely scanned. In the nonlinear regimes ($d_{\rm L}=626.4$ nm,  $C=0.024$ and $d_{\rm L}=626.4$ nm,  $C=0.095$) the resonant peaks are washed out}
\label{Retresults}
\end{figure}

When under the applied external force the quasimomentum explores the Brillouin zone, adiabatic transitions occur at
the points of ÒavoidedÓ crossings between the adjacent Bloch bands, for example, between the first and second bands in Fig. \ref{Bandstructure}. The probability
of this transition is given by the LZ  tunneling formula of \ref{landauzener}. In a first approximation, one can assume that
the adiabatic transition occurs once for each Bloch cycle with the period $T_{\rm B}$ of \ref{Blochperiod}.  Then the population of the initial band decreases exponentially with a  rate which, for the tunneling out of the ground $n=1$ band, is given by \cite{Niu96}
\begin{equation}
\Gamma_1=\nu_{\mathrm R} F_0 e^{-\frac{\pi^2V_0^2}{32F_0}}\,
\end{equation}
A plot of this tunneling rate as a function of $F_0$  in the
linear regime is shown in Fig. \ref{Retresults}. This regime is reached
either by choosing small radial dipole trap frequencies or by
releasing the BEC from the trap before the acceleration phase and
thus letting it expand. In both cases, the density and hence the
interaction energy of the BEC is reduced. Superimposed on the
overall exponential decay of $\Gamma_1/F_0$ with $F_0$, one
clearly sees the resonant tunneling peaks corresponding to $\Delta
i=3,2$. For this choice of parameters, the $\Delta i=1$
peak lay outside the region of $F$ values explored in the experiment. In order to
highlight the deviation from the LZ prediction, the dashed line represents the prediction of \ref{landauzener}.  The experimental results
are in good agreement with numerical solutions obtained by
diagonalizing the Hamiltonian of the open decaying
system represented by the continuous line.

Resonances in quantum tunneling for atomic motion within an optical lattice were previously observed by few authors. Evidence of RET is apparent in the $\Gamma$ measurements on cold atoms of~\cite{Bharucha97}. In a gray optical lattice they appear as a magnetization modulation~\cite{Teo02}. In the demonstration of the Mott insulator phase of \cite{Greiner02,GreinerPhD}  with each lattice site of Fig. \ref{WannierStarkConf} occupied by a single atom, RET occurred when  the energy difference between neighbouring lattice sites was equal to the on-site atomic interaction energy.

\section{Conclusions}
In recent years quantum-wave transport phenomena linked to Bloch oscillations and LZ tunneling in a variety of optical lattice configurations have been widely investigated.  In addition, Bloch oscillations of ultracold atoms were proposed as a tool for precision  measurements of tiny forces with a spatial resolution at the micron level~\cite{Clade05}. In \cite{Ferrari06} Bloch oscillations of ultracold  atoms were performed within a few microns from a test mass in order to  measure gravity with very large accuracy in order to test deviations from Newtonian law.  Measurements of the recoil velocity of rubidium atoms based on Bloch oscillations lead to an accurate determination of the fine structure constant~\cite{Clade06}. Bloch oscillations of ultracold fermionic atoms  have also been proposed  as a sensitive measurement of forces at the micrometer length scale,  in order to perform a local and direct measurement of the Casimir-Polder force~\cite{Carusotto05}.  The use of quantum resonant tunneling that presents a resonant dependence on the external force may improve the accuracy of those measurements.

\section{Acknowledgments}
The research work presented here relied on  the collaborative effort of M. Anderlini, M. Cristiani,  M. Jona-Lasinio, H. Lignier, J.H. M\"uller,  R. Mannella, C. Sias, Y. Singh, S. Wimberger   and A. Zenesini.
This work was supported by the European Community OLAQUI and EMALI Projects, and by MIUR-PRIN Projects.


\bibliographystyle{spphys}

\begin{thebibliography}{99.}

\bibitem{Morsch02}
O.~Morsch, E.~Arimondo, in \emph{Dynamics and Thermodynamics of Systems with
  Long-Range Interactions}, ed. by T.~Dauxois, S.~Ruffo, E.~Arimondo,
  M.~Wilkens (Springer-Verlag, 2002), p. 312

\bibitem{Bloch05a}
I.~Bloch, J. Phys. B: At. Mol. Opt. Phys. \textbf{38}, S629 (2005)

\bibitem{Bloch05b}
I.~Bloch, Nat. Phys. \textbf{1}, 23 (2005)

\bibitem{Morsch06}
O.~Morsch, M.~Oberthaler, Rev. Mod. Phys. \textbf{78}, 180 (2006)

\bibitem{Greiner02}
M.~Greiner, O.~Mandel, T.~Esslinger, T.W. H{\"a}nsch, I.~Bloch, Nature
  \textbf{415}, 39 (2002)

\bibitem{AshcroftMermin76}
N.~Ashcroft, N.D. Mermin, \emph{Solid State Physics} (International Thomson
  Publishing, New York, 1976)

\bibitem{Bloch29}
F.~Bloch, Z. Phys. \textbf{52}(7-8), 555 (1929)

\bibitem{Landau32}
L.~Landau, Phys. Z. Sowjetunion \textbf{2}, 46 (1932)

\bibitem{Zener32}
G.~Zener, Proc. R. Soc. London, Ser. A \textbf{137}, 696 (1932)

\bibitem{Wu03}
B.~Wu, Q.~Niu, New J. Phys.  \textbf{5}, 104 (2003)

\bibitem{BenDahan96}
M.B. Dahan, E.~Peik, J.~Reichel, Y.~Y.~Castin, C.~Salomon, Phys. Rev. Lett.
  \textbf{76}(24), 4508 (1996)

\bibitem{Peik97a}
E.~Peik, M.~Ben~Dahan, I.~Bouchoule, Y.~Castin, C.~Salomon, Phys. Rev. A
  \textbf{55}(4), 2989 (1997)

\bibitem{Peik97b}
E.~Peik, M.~Ben~Dahan, I.~Bouchoule, C.~Salomon, Appl. Phys. B \textbf{65}, 685
  (1997)

\bibitem{Waschke93}
C.~Waschke, H.G. Roskos, R.~Schwedler, K.~Leo, H.~Kurz, K.~K\"ohler, Phys. Rev.
  Lett. \textbf{70}(21), 3319 (1993)

\bibitem{Lyssenko97}
V.G. Lyssenko, G.~Valu\ifmmode~\check{s}\else \v{s}\fi{}is, F.~L\"oser,
  T.~Hasche, K.~Leo, M.M. Dignam, K.~K\"ohler, Phys. Rev. Lett. \textbf{79}(2),
  301 (1997)

\bibitem{Pertsch99}
T.~Pertsch, P.~Dannberg, W.~Elflein, A.~Br\"auer, F.~Lederer, Phys. Rev. Lett.
  \textbf{83}(23), 4752 (1999)

\bibitem{Morandotti99}
R.~Morandotti, U.~Peschel, J.S. Aitchison, H.S. Eisenberg, Y.~Silberberg, Phys.
  Rev. Lett. \textbf{83}(23), 4756 (1999)

\bibitem{Trompeter06}
H.~Trompeter, W.~Krolikowski, D.N. Neshev, A.S. Desyatnikov, A.A. Sukhorukov,
  Y.S. Kivshar, T.~Pertsch, U.~Peschel, F.~Lederer, Phys. Rev. Lett.
  \textbf{96}(5), 053903 (2006)

\bibitem{SanchisAlepuz07}
H.~Sanchis-Alepuz, Y.A. Kosevich, J.~S\'{a}nchez-Dehesa, Phys. Rev. Lett.
  \textbf{98}(13), 134301 (2007)

\bibitem{Morsch01}
O.~Morsch, J.H. M\"uller, M.~Cristiani, D.~Ciampini, E.~Arimondo, Phys. Rev.
  Lett. \textbf{87}(14), 140402 (2001)

\bibitem{Cristiani02}
M.~Cristiani, O.~Morsch, J.H. M\"uller, D.~Ciampini, E.~Arimondo, Phys. Rev. A
  \textbf{65}(6), 063612 (2002)

\bibitem{HeckerDenschlag02}
J.~{Hecker Denschlag}, J.~Simsarian, H.~H\"{a}ffner, C.~McKenzie, A.~Browaeys,
  D.~Cho, K.~Helmerson, S.~Rolston, W.D. Phillips, J. Phys. B: At. Mol. Opt.
  Phys. \textbf{35}(14), 3095 (2002)

\bibitem{Browaeys05}
A.~Browaeys, H.~H\"{a}ffner, C.~McKenzie, S.L. Rolston, K.~Helmerson, W.D.
  Phillips, Phys. Rev. A \textbf{72}(5), 053605 (2005)

\bibitem{Schmid06}
S.~Schmid, G.Thalhammer, K.~Winkler, F.~Lang, J.H. Denschlag, New J. Phys.
  \textbf{8}(8), 159 (2006)

\bibitem{Rubbmark81}
J.R. Rubbmark, M.M. Kash, M.G. Littman, D.~Kleppner, Phys. Rev. A
  \textbf{23}(6), 3107 (1981)

\bibitem{Bouwmeester95}
D.~Bouwmeester, N.H. Dekker, F.E.v. Dorsselaer, C.A. Schrama, P.M. Visser, J.P.
  Woerdman, Phys. Rev. A \textbf{51}(1), 646 (1995)

\bibitem{Bharucha97}
C.F. Bharucha, K.W. Madison, P.R. Morrow, S.R. Wilkinson, B.~Sundaram, M.G.
  Raizen, Phys. Rev. A \textbf{55}(2), R857 (1997)

\bibitem{Jona03}
M.~Jona-Lasinio, O.~Morsch, M.~Cristiani, N.~Malossi, J.H. M\"uller,
  E.~Courtade, M.~Anderlini, E.~Arimondo, Phys. Rev. Lett. \textbf{91}(23),
  230406 (2003). Erratum  {\it ibidem}  \textbf{93}(11),
  119903 (2004)



\bibitem{Esaki73}
L.~Esaki, in \emph{Nobel Lectures, Physics 1971-1980}, ed. by S.~Lundqvist
  (World Scientific, Singapore, 1992)

\bibitem{Niu96}
Q.~Niu, X.G. Zhao, G.A. Georgakis, M.G. Raizen, Phys. Rev. Lett.
  \textbf{76}(24), 4504 (1996)

\bibitem{Teo02}
B.K. Teo, J.R. Guest, G.~Raithel, Phys. Rev. Lett. \textbf{88}(17), 173001
  (2002)

\bibitem{GreinerPhD}
M.~Greiner, Ultracold quantum gases in three-dimensional optical lattice
  potentials.
\newblock Ph.D. thesis, Ludwig-Maximilians-Universit\"at M\"unchen (2003)

\bibitem{Clade05}
P.~Clad\'e, S.~Guellati-Kh\'elifa, C.~Schwob, F.~Nez, L.~Julien, F.~Biraben,
  Europhys. Lett. \textbf{71}(5), 730 (2005)

\bibitem{Ferrari06}
G.~Ferrari, N.~Poli, F.~Sorrentino, G.M. Tino, Phys. Rev. Lett. \textbf{97}(6),
  060402 (2006)

\bibitem{Clade06}
P.~Clad\'{e}, E.~de~Mirandes, M.~Cadoret, S.~Guellati-Kh\'{e}lifa, C.~Schwob,
  F.~Nez, L.~Julien, F.~Biraben, Phys. Rev. Lett. \textbf{96}(3), 033001 (2006)

\bibitem{Carusotto05}
I.~Carusotto, L.~Pitaevskii, S.~Stringari, G.~Modugno, M.~Inguscio, Phys. Rev.
  Lett. \textbf{95}(9), 093202 (2005)

 \end{thebibliography}
%


\printindex
\end{document}